%% file: main.tex
\documentclass[acmsmall]{acmart}

\usepackage{enumitem}
\usepackage{array}

\usepackage{hyperref}

\usepackage{tabularray}
\usepackage{multirow}
\usepackage[normalem]{ulem}
\AtBeginDocument{%
  \providecommand\BibTeX{{%
    \normalfont B\kern-0.5em{\scshape i\kern-0.25em b}\kern-0.8em\TeX}}}

\setcopyright{rightsretained}
\acmDOI{10.1145/3643776}
\acmYear{2024}
\copyrightyear{2024}
\acmSubmissionID{fse24main-p1346-p}
\acmJournal{PACMSE}
\acmVolume{1}
\acmNumber{FSE}
\acmArticle{50}
\acmMonth{7}
\received{2023-09-29}
\received[accepted]{2024-01-23}

\renewcommand\fbox{\fcolorbox{red}{white}}

\newcommand{\zc}[1]{\textcolor{red}{\textbf{ZC:} #1}}
\newcommand{\rwr}[1]{\textcolor{red}{\textbf{Reviewer:} #1}}

\newcommand{\xiaoxue}[1]{\textcolor{orange}{\textbf{Xiaoxue}: #1}}
\begin{document}

\title{Refactoring to Pythonic Idioms: A Hybrid Knowledge-Driven Approach Leveraging Large Language Models}






\author{Zejun Zhang}
\orcid{0009-0007-8877-4762}
\affiliation{%
  \institution{Australian National University}
  \city{Canberra}
  \country{Australia}
}
\affiliation{%
  \institution{CSIRO's Data61}
  \city{Canberra}
  \country{Australia}
}
\email{zejun.zhang@anu.edu.au}

\author{Zhenchang Xing}
\orcid{0000-0001-7663-1421}
\affiliation{%
  \institution{CSIRO's Data61}
  \city{Canberra}
  \country{Australia}
}
\affiliation{%
  \institution{Australian National University}
  \city{Canberra}
  \country{Australia}
}
\email{zhenchang.xing@data61.csiro.au}

\author{Xiaoxue Ren}\authornote{Corresponding author.}
\orcid{0000-0002-5526-1617}
\affiliation{%
  \institution{Zhejiang University}
  \city{Hangzhou}
  \country{China}
}
\email{xxren@zju.edu.cn}

\author{Qinghua Lu}
\orcid{0000-0002-9466-1672}
\affiliation{%
  \institution{CSIRO's Data61}
  \city{Sydney}
  \country{Australia}
}
\email{qinghua.lu@data61.csiro.au}

\author{Xiwei Xu}
\orcid{0000-0002-2273-1862}
\affiliation{%
  \institution{CSIRO's Data61}
  \city{Sydney}
  \country{Australia}
}
\email{xiwei.xu@data61.csiro.au}

\begin{abstract} 
Pythonic idioms are highly valued and widely used in the Python programming community. 
However, many Python users find it challenging to use Pythonic idioms. 
Adopting rule-based approach or LLM-only approach is not sufficient to overcome three persistent challenges of code idiomatization including code miss, wrong detection and wrong refactoring. 
Motivated by the determinism of rules and adaptability of LLMs, we propose a hybrid approach consisting of three modules. 
We not only write prompts to instruct LLMs to complete tasks, but we also invoke Analytic Rule Interfaces (ARIs) to accomplish tasks. 
The ARIs are Python code generated by prompting LLMs to generate code.
We first construct a knowledge module with three elements including ASTscenario, ASTcomponent and Condition, and prompt LLMs to generate Python code for incorporation into an ARI library for subsequent use. 
After that, for any syntax-error-free Python code, we invoke ARIs from the ARI library to extract ASTcomponent from the ASTscenario, and then filter out ASTcomponent that does not meet the condition. 
Finally, we design prompts to instruct LLMs to abstract and idiomatize code, and then invoke ARIs from the ARI library to rewrite non-idiomatic code into the idiomatic code. 
Next, we conduct a comprehensive evaluation of our approach, RIdiom, and Prompt-LLM on nine established Pythonic idioms in RIdiom. 
Our approach exhibits superior accuracy, F1-score, and recall, while maintaining precision levels comparable to RIdiom, all of which consistently exceed or come close to 90\% for each metric of each idiom.
Lastly, we extend our evaluation to encompass four new Pythonic idioms. 
Our approach consistently outperforms Prompt-LLM, achieving metrics with values consistently exceeding 90\% for accuracy, F1-score, precision, and recall.

\end{abstract}





\maketitle

\section{Introduction}
\input{intro}

\section{Motivating Examples}\label{motivation}
\input{motivation}

\section{Approach}\label{method}
\input{approach}


\section{Evaluation}\label{result}
\input{result}

\section{Discussion}\label{discussion}
\input{discuss}
\section{Related Work}\label{relatedwork}
\input{related}


\section{Conclusion and Future Work}\label{conclusion}
Refactoring non-idiomatic code with Pythonic idioms is not easy because of three challenges including code miss, wrong detection and wrong refactoring. 
Depending solely on Large Language Models (LLMs) or a rule-based approach (RIdiom) has its limitations in addressing these challenges. 
To alleviate the challenges, we propose a hybrid approach based on LLMs to refactor non-idiomatic code with Pythonic idioms. 
In detail, we first extract three elements of a Pythonic idiom and create an API library by prompting LLMs to generate code. 
We then invoke the APIs to extract non-idiomatic code in the extraction module. 
Finally, we prompt LLMs to abstract code, idiomatize the abstract code and then invoke APIs to rewrite non-idiomatic code into idiomatic code. 
The results of our experiments, conducted on nine Pythonic idioms from RIdiom~\cite{zhang2023ridiom}, as well as four new Pythonic idioms~\cite{Zen_Your_Python} not covered by RIdiom, demonstrate high levels of accuracy, F1-score, precision, and recall. 
This substantiates the effectiveness and scalability of our proposed approach. 
In the future, we will keep improving our approach and extend our approach to accommodate Python code with syntax errors.
Besides, we plan to offer explanations regarding the impacts, such as enhanced readability and performance, that result from refactoring code into idiomatic Python. 
\section{Data Availability}\label{data}
Our replication package can be found here~\cite{Replication_Package}. 
\bibliographystyle{ACM-Reference-Format}

\bibliography{sample-base}




\end{document}

%% file: intro.tex
Pythonic idioms refer to programming practices and coding conventions that align with the core philosophy and style of the Python programming language~\cite{almaEffective_Python,alexandru2018usage,Zen_Your_Python,zhang2022making}. 
RIdiom~\cite{zhang2022making,zhang2023ridiom} identified nine Pythonic idioms by comparing syntax differences between Python and Java. 
Farooq et al.~\cite{Zen_Your_Python} conducted a literature review to identify and explore the usage of twenty-seven Pythonic idioms. 
An example of a Pythonic idiom is the chain-comparison idiom, which allows comparing multiple variables in one comparison operation, such as ``\textsf{-size <= x.indices(size)[0] <= size}''. 
The Python community continually strives to design and improve them to achieve code conciseness and improved performance~\cite{pep,alexandru2018usage}. 
For the above example of chain-comparison, 
in contrast to the non-idiomatic equivalent, ``\textsf{-size <= x.indices(size)[0] and x.indices(size)[0] <= size}'', the chain-comparison  simplifies code and improves performance. 
Given these benefits, the community and renowned Python developers actively promote the widespread adoption of Pythonic idioms~\cite{almaEffective_Python,beazley2013python,bader2017python, knupp2013writing,hettinger2013transforming,antipattern_python}.
However, previous studies~\cite{zhang2022making,alexandru2018usage} have indicated that Python users often be unaware of Pythonic idioms or unsure of how to correctly use Pythonic idioms, as Pythonic idioms are scattered across various materials, and are known for their versatile nature~\cite{almaEffective_Python,beazley2013python,bader2017python,antipattern_python}. 
For example, chain-comparison idiom also supports ``\textsf{in}'' operator (e.g., ``\textsf{line <= r[1] in rlist}''),  yet this usage often goes unnoticed by many Python users. 
To help Python users use Pythonic idioms, refactoring non-idiomatic code with Pythonic idioms emerges as a promising solution~\cite{phan2020teddy,zhang2022making}. 
Through pilot studies, we find the task faces three challenges including code miss, wrong detection and wrong refactoring because of the versatile nature of Pythonic idioms. 
Code miss refers to \textbf{missing non-idiomatic code that can be refactored with Pythonic idioms}. 
For example, RIdiom misses a For statement code that can be refactored with set comprehension, as shown in code \textcircled{1} of Figure~\ref{motiv_example}. 
Wrong detection refers to \textbf{misidentifying non-refactorable non-idiomatic code with Pythonic
idioms as refactorable}. 
For example, Prompt-LLM wrongly thinks a For statement without adding elements to a set as refactorable with set comprehension, as shown in code \textcircled{5} of Figure~\ref{motiv_example}. 
Wrong refactoring refers to \textbf{giving wrong idiomatic code for refactorable non-idiomatic code with Pythonic
idioms}. 
For example, Prompt-LLM does not correctly refactor code ``\textsf{0< y\_int <h\_i and w\_i < 0}'' with chain comparison, the corresponding idiomatic code is ``\textsf{w\_i < 0 < y\_int < h\_i}'' as shown in code \textcircled{3} of Figure~\ref{motiv_example}. 


A most related work, RIdiom~\cite{zhang2023ridiom}, employs a rule-based approach to establish detection rules and
refactoring procedures to refactor the non-idiomatic code into idiomatic code for nine
Pythonic idioms. 
However, it is noteworthy that when confronted with intricate instances of non-idiomatic code, the reliance on pre-defined, inflexible rules cannot overcome the above three challenges (see RIdiom examples of Figure~\ref{motiv_example} in Section~\ref{motivation}). 
Even when identified, formulating rules to refactor it into idiomatic code is challenging. 
On the other hand, in light of the success of Large Language Models (LLMs), users can simply describe natural language prompts to instruct LLMs to perform specific software engineering tasks, such as code generation~\cite{dong2023selfcollaboration,code_gen_eval_tianyi,codex,fried2023incoder,Nijkamp2022CodeGenAO} and program synthesis~\cite{Prompt_Tune_qinghuang,huang2023ai,peng2023generative,huang2023chain}. 
It inspires us to explore LLMs for code idiomatization, wherein we observe their powerful ability in certain scenarios, e.g., code \textcircled{1} of Figure~\ref{motiv_example} can be correctly refactored with set comprehension by LLMs. 
However, without knowledge guiding, LLMs may make obvious mistakes that can be easily avoided using rules because of the inherent randomness and black boxes of LLMs (see Prompt-LLM examples of Figure~\ref{motiv_example} in Section~\ref{motivation}). 
This observation underscores the insufficiency of relying solely on rule-based approach or LLMs. 
It motivates us to propose a hybrid approach that combines the determinism inherent in rule-based approach with the adaptability offered by LLMs. 
Specifically, our hybrid approach comprises three core modules: \textbf{a knowledge module}, \textbf{an extraction module}, and \textbf{an idiomatization module}. 
For each module, we write prompts to instruct LLMs to complete tasks or invokes Analytic Rule Interfaces (ARIs) to complete tasks. 
ARIs are Python code generated by prompting LLMs to generate code.
The knowledge module is to construct a knowledge base consisting of three elements of non-idiomatic code of thirteen Pythonic idioms and an ARI library. 
The three elements are ASTscenario (the usage scenario of non-idiomatic code), ASTcomponent (the composition of non-idiomatic code) and Condition (the condition that refactorable non-idiomatic code must meet). 
The ARI library consists of ARIs to extract three elements and auxiliary ARIs to rewrite non-idiomatic code into idiomatic code. 
In the extraction module, for any syntax-error-free Python code, we invoke ARIs from the ARI library to extract ASTscenario and ASTcomponent that satisfies the condition, which will be input into the idiomatization module. 
The idiomatization module consists of three steps: abstracting code, idiomatizing code and rewriting code.  
We first abstractly represent the code of ASTcomponent by prompting LLMs. 
We then idiomatize the abstract code through LLM prompts, producing an abstract idiomatic code. 
Finally, we utilize ARIs to rewrite the non-idiomatic code into idiomatic code by using the abstract idiomatic code. 

We conduct two experiments to evaluate the effectiveness and scalability of our approach. 
For effectiveness, we examine nine Pythonic idioms identified by RIdiom~\cite{zhang2023ridiom}. 
To determine a complete, correct and unbias benchmark, 
we randomly sample methods from the methods of each Pythonic idiom in RIdiom~\cite{zhang2023ridiom}. 
We independently run our approach, RIdiom and Prompt-LLM for the sampled methods to generate code pairs, and invite external workers to verify the correctness of code pairs by each approach manually. 
Then two authors and external workers discuss and resolve the inconsistencies. 
The metrics of accuracy, F1-score, precision, and recall were employed for evaluating results. 
The results demonstrate our approach achieves the best performance in accuracy, F1-score and recall compared to RIdiom and Prompt-LLM and achieves comparable precision with RIdiom. 
To evaluate the scalability of our approach, we choose four new Pythonic idioms not covered by RIdiom~\cite{zhang2023ridiom}. 
To avoid the bias of the benchmark, we randomly sample 600 methods from all methods in RIdiom~\cite{zhang2023ridiom}. 
To ensure the correctness and completeness of our approach, we following the same process in Section~\ref{rq1_approa}.
Since RIdiom does not support idiomatization for the four Pythonic idioms. 
We do not run RIdiom on the new four idioms. 
Our approach consistently outperformed in accuracy, F1-score, precision, and recall, all surpassing 90\% for each Pythonic idiom, which shows that our approach can be effectively extend to new Pythonic idioms. 

In summary, the contributions of this paper are as follows:
\vspace{-1mm}
\begin{itemize}[leftmargin=*]
    \item This is the first work to exploit LLMs into code idiomatization with Pythonic idioms, paving the way for new opportunities in code idiomatization. 
    \item We propose a hybrid knowlege-driven approach with ARIs and prompts based on LLMs to refactor non-idiomatic code into idiomatic code with Pythonic idioms.
    \item We conduct experiments on both established and new Pythonic idioms. 
    The high accuracy, F1-score, precision and recall verify the effectiveness and scalability of our approach.  
    We provide a replication package~\cite{Replication_Package} for future studies. 
\end{itemize}

%% file: motivation.tex
\begin{figure}
  \centering
    \includegraphics[width=5.4in]{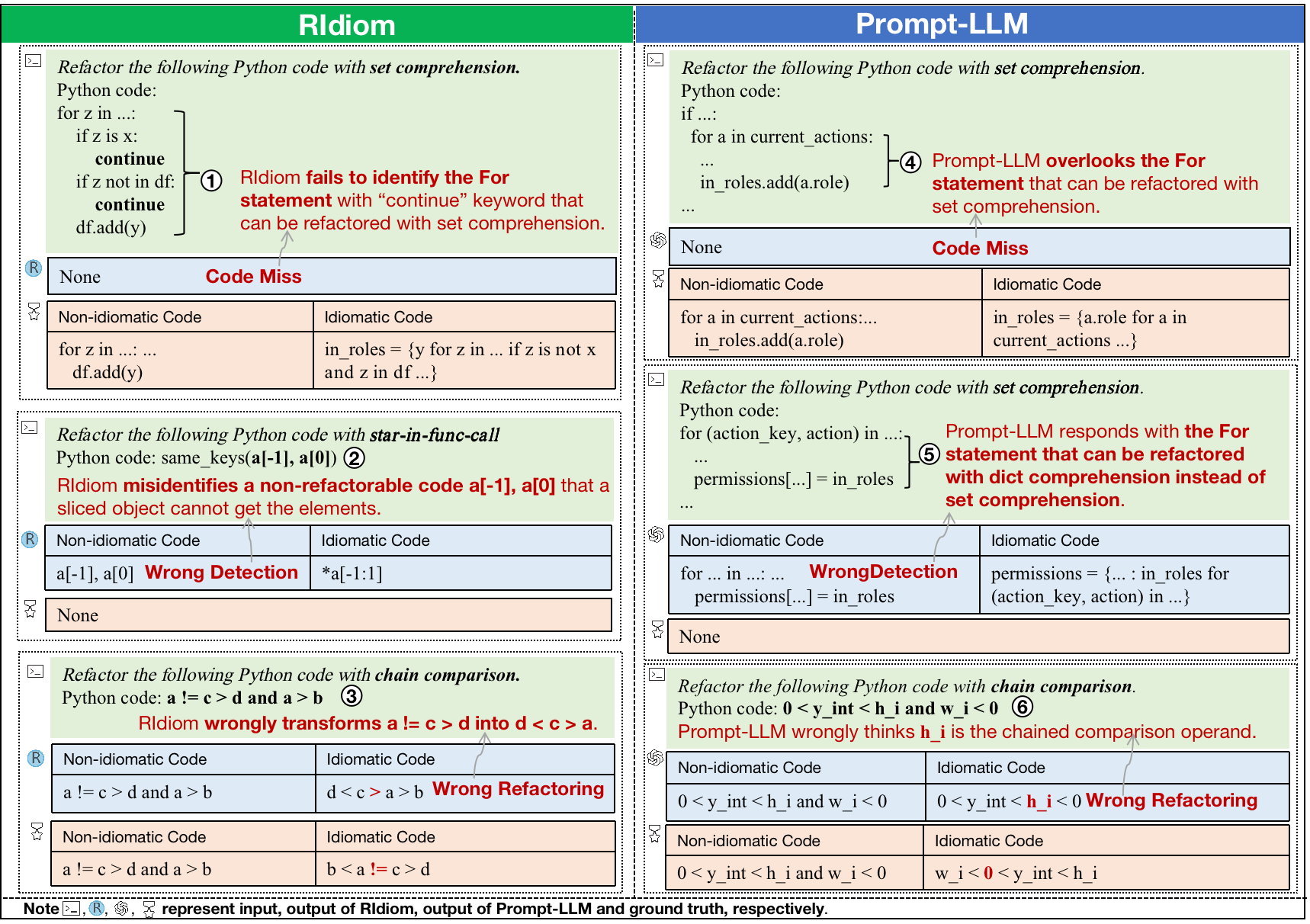}
    \caption{Motivating examples}
    \label{motiv_example} 
    \vspace{-0.3cm}
\end{figure}

Although using Pythonic idioms can improve the conciseness and performance~\cite{zhang2022making,alexandru2018usage,faster_slower,leelaprute2022does}, refactoring non-idiomatic Python code with Pythonic idioms for a given Python code is not easy. 
RIdiom~\cite{zhang2023ridiom} is the state-of-the-art rule-based approach that formulates detection and refactoring rules to automatically refactor non-idiomatic code into idiomatic code for nine Pythonic idioms. 
Recently, large language models (LLMs) have achieved great success in various software engineering tasks~\cite{brown2020language, code_generation_exception, openai2023gpt4,feng2023prompting,peng2023generative,huang2023ai}. 
LLMs can directly complete various tasks by receiving natural language prompts as input, a process we refer to as Prompt-LLM. 
To explore challenges encountered in this endeavor for the two approaches, we randomly collect ten Python methods crawled by RIdiom~\cite{zhang2023ridiom}. 
The methods may contain several non-idiomatic code that can be refactored with a Pythonic idiom.  
Next, we apply RIdiom and Prompt-LLM to each Python method for nine Pythonic idioms from RIdiom. 
Two authors collaborate to check results of RIdiom and Prompt-LLM and then identify challenges encountered by the two approaches. 
We summarize three challenges that are described as follows. 



\noindent \textbf{(1) Code Miss: miss non-idiomatic code that can be refactored with Pythonic idioms.} 
The code written by Python users comes in various styles, the code may contain several refactorable non-idiomatic code with a Pythonic idiom and the form of non-idiomatic code may be diverse. 
Missing the refactorable non-idiomatic code can lead to redundant code and performance degradation. 
Unfortunately, code missing is common  in the two approaches. 
On the one hand, given the diverse and intricate nature of non-idiomatic code patterns, some instances may pose challenges that surpass the capabilities of straightforward rule-based identification. 
For example, code \textcircled{1} of the RIdiom column of Figure~\ref{motiv_example} is a ``\textsf{for}'' statement with two ``\textsf{continue}'' statements that can be refactored with set-comprehension. 
Since set-comprehension does not support ``\textsf{continue}'' keyword, RIdiom wrongly assumes the code cannot be refactored. 
Actually, we can change ``\textsf{z is x}'' and ``\textsf{z not in df}'' into ``\textsf{z is not x}'' and ``\textsf{z in df}'', and then we use the ``\textsf{and}'' to connect the two conditions to remove continue statements. 
On the other hand,
in a codebase, instances of non-idiomatic code are distributed throughout, necessitating a comprehensive scan of the entire codebase to identify such occurrences. 
Unlike rule-based programs that deterministically scan Python code from start to end, LLMs operate as black boxes. 
This non-deterministic nature can inadvertently lead to the oversight of refactorable non-idiomatic code~\cite{alexandru2018usage,Zen_Your_Python,faster_slower}. 
For example, code \textcircled{4} of the Prompt-LLM column in Figure~\ref{motiv_example} shows LLMs miss a ``\textsf{for}'' statement that can be refactored with set-comprehension. 

\noindent \textbf{(2) Wrong Detection: misidentify non-refactorable non-idiomatic code with Pythonic idioms as refactorable,} which can lead to misunderstandings among Python users regarding Pythonic idioms and potentially introducing bugs into the codebase.  
Although not common in RIdiom, it should not be ignored. 
The detection rules of RIdiom are human-defined, and developers may overlook the nuances of code structures and Python syntax semantics, leading to false discoveries. 
For example, for \textcircled{2} of  Figure~\ref{motiv_example}, RIdiom determines ``-1, 0'' is an arithmetic sequence, so it wrongly thinks that ``\textsf{a[-1], a[0]}'' can be obtained by a sliced object ``\textsf{a[-1:1]}''.  
However, since Python list grows linearly and is not cyclic, as such slicing does not wrap (from end back to start going forward) as we expect, the ``\textsf{a[-1:1]}'' actually is empty. 
On the other hand, although LLM has powerful abilities, its flexibility and adaptability usually cause inappropriate or off-topic response. 
For example, code \textcircled{5} of the Prompt-LLM column in Figure~\ref{motiv_example} shows that the Prompt-LLM wrongly classifies a ``\textsf{for}'' statement that can be refactored with dict-comprehension as refactorable non-idiomatic code with set-comprehension. 
Actually, we can add a condition to check whether the for statement has an ``\textsf{add}'' function call to filter out the wrong detection. 
For another example, non-idiomatic code of chain comparison should have two comparison operations. However, Prompt-LLM often mistakenly suggests refactoring a single comparison operation with chain comparison, even though it cannot be refactored in this way. 
For instance, when encountering a single comparison operation like ``\textsf{start is not None}'', Prompt-LLM wrongly assumes it can be refactored using chain comparison.



\noindent \textbf{(3) Wrong Refactoring: give wrong idiomatic code for refactorable non-idiomatic code with Pythonic
idioms.} 
It occurs in identifying refactorable non-idiomatic code but wrongly refactoring it, 
resulting in inconsistency in code behavior before and after refactoring.  
The diversity and complexity of such non-idiomatic code make both the rule-based approach and the Prompt-LLM approach prone to errors. 
For example, for the chain-comparison, to chain two comparison operations into one comparison operation, we need to reverse compare operands for each comparison operation and consider if we need to change the comparison operation. 
When one comparison operation has more than one comparison operation, it is more likely to make mistakes. 
For example, code \textcircled{3} of the RIdiom column in Figure~\ref{motiv_example} shows that RIdiom wrongly transforms ``\textsf{a != c > d}'' into ``\textsf{d < c > a}''. 
Directly using LLMs may make unexpected mistakes. 
For example, code \textcircled{6} of Prompt-LLM column of Figure~\ref{motiv_example} shows that Prompt-LLM assumes that ``\textsf{h\_i}'' is the chained comparison operand and then wrongly refactors it into ``\textsf{0 < y\_int < h\_i < 0}''. 

The three challenges shown in  Figure~\ref{motiv_example} indicate that the rule-based approach, while deterministic, may still fall short in identifying all refactorable non-idiomatic code instances, especially those that are inherently complex or difficult to address through formulating rules (e.g., \textcircled{1} of Figure~\ref{motiv_example}). 
Conversely, relying solely on the flexibility and adaptability of LLMs without knowledge guidance can lead LLMs to make obvious mistakes. 
For example, refactorable non-idiomatic code with set comprehension should contain an ``\textsf{add}'' function call. 
Regrettably, code \textcircled{5} of Figure~\ref{motiv_example} lacks this function call.  
The absence of this contextual knowledge leads LLMs to misidentify it can be refactored with set comprehension. 
Therefore, a judicious approach emerges: initially employing code to handle deterministic and simple tasks, and then leveraging LLMs to tackle the more challenging refactoring endeavors where the rule-based approach may struggle. 
This hybrid approach stands poised to offer a comprehensive and effective solution. 

\vspace{1mm}
\noindent\fbox{\begin{minipage}{5.4in} \emph{ Make Python code idiomatic encounters three challenges, including code miss, wrong detection and wrong refactoring.  
Adopting only one approach (rule-based approach or LLMs) alone is not sufficient.  
} \end{minipage}}

%% file: approach.tex
\begin{figure}
\vspace{-0.2cm}
  \centering
   \includegraphics[width=5.3in]{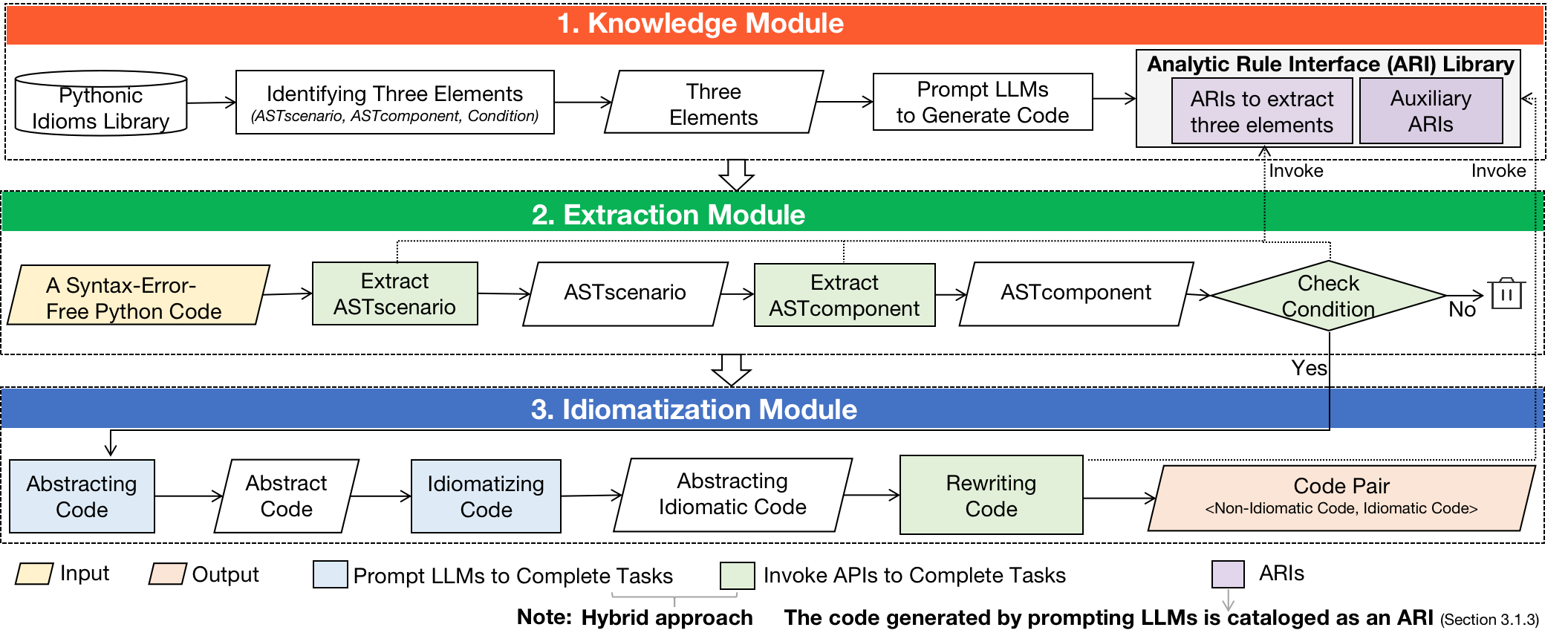}
    \caption{Approach overview}
    \label{fig:framework_chain_compare} 
    \vspace{-0.3cm}
\end{figure}

Inspired by the motivating examples in Section~\ref{motivation}, we propose a hybrid approach based on LLMs to refactor non-idiomatic code with Pythonic idioms. 
Figure~\ref{fig:framework_chain_compare} shows the approach overview. 
We first construct knowledge base of non-idiomatic code of Pythonic idioms, which consists of three elements: ASTscenario, ASTcomponent and Condition, and an ARI library consisting of ARIs to extract the three elements and ARIs to rewrite code.  
After that, for a given Python code, we first call ARIs to extract its ASTcomponent from the ASTscenario, and then filter out ASTcomponent that does not meet the condition. 
Then we input the ASTcomponent and ASTscenario from extraction module into the idiomatization module.  
The idiomatization module consists of three steps: abstracting code, idiomatizing code and rewriting code.
To reduce the pressure on LLMs to idiomatize code, we first abstract code by abstracting expression of the code corresponding ASTcomponent. 
And then we write prompts to make LLMs idiomatize the abstract code. 
After we get the abstract idiomatic code, we need to replace the abstract expression with the original expressions, and then rewrite the non-idiomatic code with the idiomatic code. 
Since the rewriting operations are simple, we call ARIs from the Auxiliary ARIs to complete.

\subsection{Knowledge Module}
As investigated in Section~\ref{motivation}, lacking specific knowledge, directly using LLMs to find a non-idiomatic code with a Pythonic idiom from a given Python code is like finding a needle in a haystack. 
For example, the non-idiomatic code of set-comprehension should have a For node containing an add function call. 
Without the knowledge, Prompt-LLM misses one For node as shown in \textcircled{4} of Figure~\ref{motiv_example}, and misidentifies a For node without the add function call as non-idiomatic code of set-comprehension as shown in \textcircled{5} of Figure~\ref{motiv_example}. 
We observe the non-idiomatic code that can be refactored with Pythonic idioms has deterministic knowledge. 
Therefore, we can construct a knowledge base to boost the ability of LLMs. 

\subsubsection{Pythonic idioms library} \label{idiom_library}

\begin{table}[]
\scriptsize
\caption{Pythonic Idioms Library for Thirteen Pythonic Idioms}

\vspace{-0.2cm}
  \label{tab:four_idiom_example}
\centering
\begin{tabular}{|@{}l@{}|l@{}|l@{}|l@{}|l@{}|}
\hline
\multicolumn{1}{|c|}{Source} & \multicolumn{1}{c|}{Idiom} & \multicolumn{1}{c|}{Explanation} & \multicolumn{1}{c|}{Non-Idiomatic Code} & \multicolumn{1}{c|}{Idiomatic Code} \\ \hline
\multirow{3}{*}{\begin{tabular}[c]{@{}l@{}}RIdiom~\cite{zhang2023ridiom} \\ , Farooq\\ et al.~\cite{Zen_Your_Python}\end{tabular}} & \begin{tabular}[c]{@{}l@{}}list/set/dict\\-comprehension\end{tabular} & \begin{tabular}[c]{@{}l@{}}Use one line to append elements\\ to an iterable\end{tabular} &\begin{tabular}[c]{@{}l@{}}\textbf{new\_cols = []}\\ \textbf{for col in old\_cols:}\\
    \ \ \textbf{new\_cols.append(col + postfix)}\end{tabular} & \begin{tabular}[c]{@{}l@{}}\textbf{new\_cols = [col + postfix} \\\textbf{for col in old\_cols]}\end{tabular}  \\ \cline{2-5} 
 & \begin{tabular}[c]{@{}l@{}}chain-\\comparison\end{tabular} & \begin{tabular}[c]{@{}l@{}}Chain mutliple comparison expressions\\ into one comparison expression~\cite{zhang2022making}\end{tabular} &\begin{tabular}[c]{@{}l@{}}\textbf{a > b and a < 1}\end{tabular}  &\begin{tabular}[c]{@{}l@{}}\textbf{b < a < 1}\end{tabular}  \\ \cline{2-5} 
 & truth-test &\begin{tabular}[c]{@{}l@{}}Directly check the ``truthiness'' of\\ an object\end{tabular} & \begin{tabular}[c]{@{}l@{}}\textbf{embedding\_dim \% 2 == 0}\end{tabular} & \begin{tabular}[c]{@{}l@{}}\textbf{not embedding\_dim \% 2}\end{tabular} \\ \hline
\multirow{4}{*}{RIdiom~\cite{zhang2023ridiom}} & loop-else &\begin{tabular}[c]{@{}l@{}}A loop statement has an else clause\end{tabular}  & \begin{tabular}[c]{@{}l@{}}while attempt < 3:\\ \ \  ...\\ \ \  if body is not None:\\
      \ \ \ \  break\\ \textbf{if body is None:} \\\ \  ...\end{tabular} & \begin{tabular}[c]{@{}l@{}}while attempt < 3:\\ \ \ ...\\ \ \  if body is not None:\\
       \ \ \ \ break\\ \textbf{else}: \\\ \ ...\end{tabular} \\ \cline{2-5} 
 & assign-multi-tar &\begin{tabular}[c]{@{}l@{}}Assign multiple values to multiple\\ variables in an assign statement\end{tabular}  & \begin{tabular}[c]{@{}l@{}}\textbf{self.\_ad = device}\\
\textbf{self.\_sl4a\_client = None} \end{tabular}& \begin{tabular}[c]{@{}l@{}}\textbf{self.\_ad, self.\_sl4a\_client =} \\\textbf{device, None} \end{tabular} \\ \cline{2-5} 
 & for-multi-tar & \begin{tabular}[c]{@{}l@{}}Unpack the iterated target of a for\\ statement\end{tabular} & \begin{tabular}[c]{@{}l@{}}for \textbf{sample} in family.samples:\\
    \ \ if \textbf{sample[0]} > 2: \\\ \  \ \ ...\end{tabular} &\begin{tabular}[c]{@{}l@{}}for \textbf{e0, *e} in family.samples:\\
    \ \ if \textbf{e0} > 2: \\\ \  \ \ ...\end{tabular}  \\ \cline{2-5} 
 & star-in-func-call & \begin{tabular}[c]{@{}l@{}}Unpack an iterable to the positional\\ arguments in a function call\end{tabular} & \begin{tabular}[c]{@{}l@{}}nn.Linear(\textbf{gate\_channels[i]}\\\textbf{, gate\_channels[i+1])}\end{tabular} & \begin{tabular}[c]{@{}l@{}}nn.Linear(\textbf{*gate\_channels[i:}\\\textbf{i + 2]})\end{tabular} \\ \hline
\multirow{4}{*}{\begin{tabular}[c]{@{}l@{}}Farooq\\ et al.~\cite{Zen_Your_Python}\end{tabular}} &\begin{tabular}[c]{@{}l@{}}with\end{tabular}       & \begin{tabular}[c]{@{}l@{}} Automatically close a file after\\ it has been
opened
\end{tabular} & 
\begin{tabular}[c]{@{}l@{}}  
bamfiles = [x.strip() for\\ x in \textbf{open(bamfile)}]
\end{tabular}             & 
\begin{tabular}[c]{@{}l@{}}
\textbf{with open(bamfile) as f:}\\
 \ \ \ \       bamfiles = [x.strip() for x in f]\\
\end{tabular}  \\ \cline{2-5} 
 & enumerate & \begin{tabular}[c]{@{}l@{}} Return a tuple containing a count\\ (from start which defaults to 0) and\\ 
the values obtained from iterating\\ over iterable.\end{tabular}& 
\begin{tabular}[c]{@{}l@{}}
for \textbf{i} in \textbf{range(len(text))}:\\
\ \      \textbf{w = text[i]}\\
\ \      if w in token2id:\\
\ \  \ \          R[i] = token2id[w]\\

\end{tabular}             & 
\begin{tabular}[c]{@{}l@{}}
for \textbf{(i, w)} in \textbf{enumerate(text)}:\\
\ \       if w in token2id:\\
 \ \   \ \          R[i] = token2id[w]
\end{tabular}\\ \cline{2-5} 
 & \begin{tabular}[c]{@{}l@{}}chain-ass-\\same-value\end{tabular} & \begin{tabular}[c]{@{}l@{}}Assign a value to multiple \\variables.\end{tabular} & \begin{tabular}[c]{@{}l@{}}
\textbf{global\_draw\_name = None} \\
\textbf{\_test\_name = None} \end{tabular}           & 
\begin{tabular}[c]{@{}l@{}}
\textbf{global\_draw\_name =} \\\textbf{\_test\_name =None}
\end{tabular}  \\ \cline{2-5} 
 & fstring &  \begin{tabular}[c]{@{}l@{}} Dynamically combine data from\\ variables and other data structures\\ into a readable string output.\end{tabular} & \begin{tabular}[c]{@{}l@{}}
log.info(`\textbf{sample\_num\_list is \%s}'\\ 
\% \textbf{repr(self.sample\_num\_list))}\end{tabular}             &
\begin{tabular}[c]{@{}l@{}}
log.info(f`\textbf{sample\_num\_list is} \\\textbf{{repr(self.sample\_num\_list)}}')
\end{tabular} \\ \hline
\end{tabular}
\vspace{-3mm}
\end{table}
Pythonic idioms are highly valued by developers~\cite{programming_idioms,hettinger2013transforming,knupp2013writing}, many studies summarize Pythonic idioms and research their usage ~\cite{alexandru2018usage,Zen_Your_Python,merchantepython,zhang2022making}. 
RIdiom~\cite{zhang2023ridiom} identified nine Pythonic idioms by comparing the syntax difference between Python and Java. 
The nine Pythonic idioms are list/set/dict-comprehension, chain-comparison, truth-test, loop-else, assign-multi-targets, for-multi-targets and star-in-func-call. 
State-of-the-art of research~\cite{Zen_Your_Python} based on a literature review identified a total of 27 detectable idioms, of which five (list/set/dict-comprehension, chain-comparison, truth-test) overlap with those defined by RIdiom. 
After excluding infrequently used idioms or those with rarely corresponding non-idiomatic Python code~\footnote{For example, @staticmethod, assert and etc. are common syntax in programming languages, a few python developers use other syntax alone to achieve the same functionality without these idioms.}, four idioms remain: with, enumerate, fstring, and chain-assign-same-value. 
This culminates in a total of 13 Pythonic idioms. 
Table~\ref{tab:four_idiom_example} gives the explanation and code examples of the 13 Pythonic idioms. 


\subsubsection{Three elements of non-idiomatic code of Pythonic idioms} 

We construct the knowledge base of non-idiomatic code of Pythonic idioms as triples of <element, relation, element>. 
It comprises three fundamental elements: ASTscenario, ASTcomponent, and Condition. 
Each element focuses on a unique aspect: ASTscenario represents usage scenarios for non-idiomatic code linked to a Pythonic idiom. 
ASTcomponent defines the composition of such code, and Condition outlines necessary conditions for the ASTcomponents. 
There exist two relationships between these elements: 
the ASTcomponent relies on ASTscenario, and the ASTcomponent adheres to the specified conditions to be considered as non-idiomatic code of Pythonic idioms. 
Table~\ref{tab:three_ele_idioms} shows the three elements of non-idiomatic code of thirteen Pythonic idioms.
The details are as follows:

\noindent \textbf{ASTscenario:} A non-idiomatic code associated with a Pythonic idiom may have restrictions on usage scenarios, corresponding to a distinct Abstract Syntax Tree (AST) node, referred to as ASTscenario.
For example, chain-comparison idiom can chain two comparison operations using the ``and'' operator into one comparison. 
So the ASTscenario is a BoolOP node whose op is ``and'' as shown in Table~\ref{tab:three_ele_idioms}. 
For another example, for the list comprehension idiom in Table~\ref{tab:three_ele_idioms}, it allows the addition of elements to an object in just one line, as opposed to using a for statement. 
Since the for statement has no restrictions on the usage scenario, it does not possess an associated ASTscenario.

\noindent \textbf{ASTcomponent}: A non-idiomatic code associated with a Pythonic idiom has a deterministic composition, corresponding to few AST nodes, referred to as ASTcomponent. 
It serves as a pivotal entity in discerning and addressing non-idiomatic code patterns.
Taking the chain-comparison idiom in Table~\ref{tab:three_ele_idioms} as an example, its ASTcomponent comprises two Compare nodes within a BoolOP node. 
These Compare nodes form essential elements of the non-idiomatic code pattern. 
For another example, for the list comprehension idiom in Table~\ref{tab:three_ele_idioms}, which involves appending elements to an object in a for statement, its ASTcomponent is a For node and an Assign node. 

\noindent \textbf{Condition:} A non-idiomatic code associated with a Pythonic idiom may entail specific conditions, referred to as Condition for its ASTcomponent. 
It serves as a guiding principle for identification of refactorable ASTcomponent and avoid LLMs from idiomatizing non-refactorable ones that does not meet the specified conditions. 
For example, for the chain-comparison idiom in Table~\ref{tab:three_ele_idioms}, the condition stipulates that the compare operands of the two Compare nodes must intersect. 
For another example, for the list-comprehension in Table~\ref{tab:three_ele_idioms}, its non-idiomatic code is to append elements to a list, so the For node of ASTcomponent should has a ``append'' function call whose function name is the assigned variable of the Assign node. 

\noindent \textbf{Relation}: 
There are two relationships between the three elements. 
The ASTcomponent is depend on the ASTscenario, the ASTcomponent should satisfy the Condition. 
For example, for the chain-comparison idiom, its AST component (two Compare nodes) is depend on the ASTscenario (BoolOp node whose op is ``and''), and its ASTcomponent satisfies the Condition (Compare operands of the two Compare nodes intersect). 
These relationships establish a clear framework for identifying non-idiomatic code. 
We elaborate it in Section~\ref{extract_module_with_APIs}.
\begin{table}[]
\scriptsize
\caption{Three Elements of Non-Idiomatic Code of Thirteen Pythonic Idioms}
\vspace{-0.3cm}
  \label{tab:three_ele_idioms}
\centering
\begin{tabular}{|l|l|l|l|}
\hline
\multicolumn{1}{|c|}{Idiom} & \multicolumn{1}{c|}{ASTscenario} & \multicolumn{1}{c|}{ASTcomponent}              & \multicolumn{1}{c|}{Condition}                                                                                                                                                                                                                                 \\ \hline
list/set-comprehension       & \multicolumn{1}{c|}{---}                              & \begin{tabular}[c]{@{}l@{}}A For node\\ An Assign node\end{tabular}                     & \begin{tabular}[c]{@{}l@{}}1. The For node has ``append/add'' function call\\
2. The function name of ``append/add'' function call\\ is the assigned variable of the Assign node\\
\end{tabular}                                                   \\ \hline
dict-comprehension           & \multicolumn{1}{c|}{---}                              & \begin{tabular}[c]{@{}l@{}}A For node
\\An Assign node 
\end{tabular}             & \begin{tabular}[c]{@{}l@{}}1. The For node has an assign statement whose\\ assigned variable is a Subscript node\\
2. The value of the Subscript node\\ is the assigned variable of the Assign node\\ 
\end{tabular} \\ \hline
chain-comparison             & \begin{tabular}[c]{@{}l@{}}A BoolOP node \\whose op is ``and'' \end{tabular}   & Two Compare nodes                              & \begin{tabular}[c]{@{}l@{}}1. Compare operands of the two Compare nodes\\ intersect \end{tabular}                                                                                                                                                                                                           \\ \hline
truth-test                   & A test-type node                   & A Compare node                                 & \begin{tabular}[c]{@{}l@{}}1. The op of  the Compare node is “== “or “!=”\\ 2. The one comparison operand is belong to EmptySet\end{tabular}                                                                                                                   \\ \hline
loop-else                    & \multicolumn{1}{c|}{---}                              & \begin{tabular}[c]{@{}l@{}}A For/While node\\An If node\end{tabular}                   & \begin{tabular}[c]{@{}l@{}}1. The For/While node has break statements\\2. The If node is the next statement of For node\end{tabular}                                                                                                                                                                         \\ \hline
assign-multi-tar             & \multicolumn{1}{c|}{---}                              & Consecutive Assign nodes                       & \multicolumn{1}{c|}{---}                                                                                                                                                                                                                                                            \\ \hline
for-multi-tar                & \multicolumn{1}{c|}{---}                              & A For node                                     & \begin{tabular}[c]{@{}l@{}}1. The body of the For node has a Subscript node\\2. The value of the Subscript node is the iterated\\ variable of the For node     \end{tabular}                                                                                                                                           \\ \hline
star-in-func-call            & A Call node                        & \begin{tabular}[c]{@{}l@{}}Consecutive Subscript nodes \end{tabular} & 1. The values of Subscript nodes are the same                                                                                                                                                                                                                                                                  \\ \hline
with                         & \multicolumn{1}{c|}{---}                              & A Call node                                    & 1. The function name of the Call node is “open”                                                                                                                                                                                                                \\ \hline
enumerate                    & \multicolumn{1}{c|}{---}                              & A For node                                     & \begin{tabular}[c]{@{}l@{}}1. The iterated object is not a function call whose\\ function name is ``enumerate'' \end{tabular}                                                                                                                                                                              \\ \hline
chain-ass-same-value         & \multicolumn{1}{c|}{---}                              & Consecutive Assign nodes                       & 1. The values of consecutive Assign nodes are the same                                                                                                                                                                                                         \\ \hline
fstring                      & \multicolumn{1}{c|}{---}                              & A BinOP node                                   & 1. The op of the BinOp node is ``\%''                                                                                                                                                                                                                            \\ \hline
\end{tabular}
\vspace{-3mm}
\end{table}

\subsubsection{ARI library}\label{APIs} 

\begin{figure}
\vspace{-0.1cm}
  \centering
    \setlength{\abovecaptionskip}{2mm}
    \includegraphics[width=5.4in]{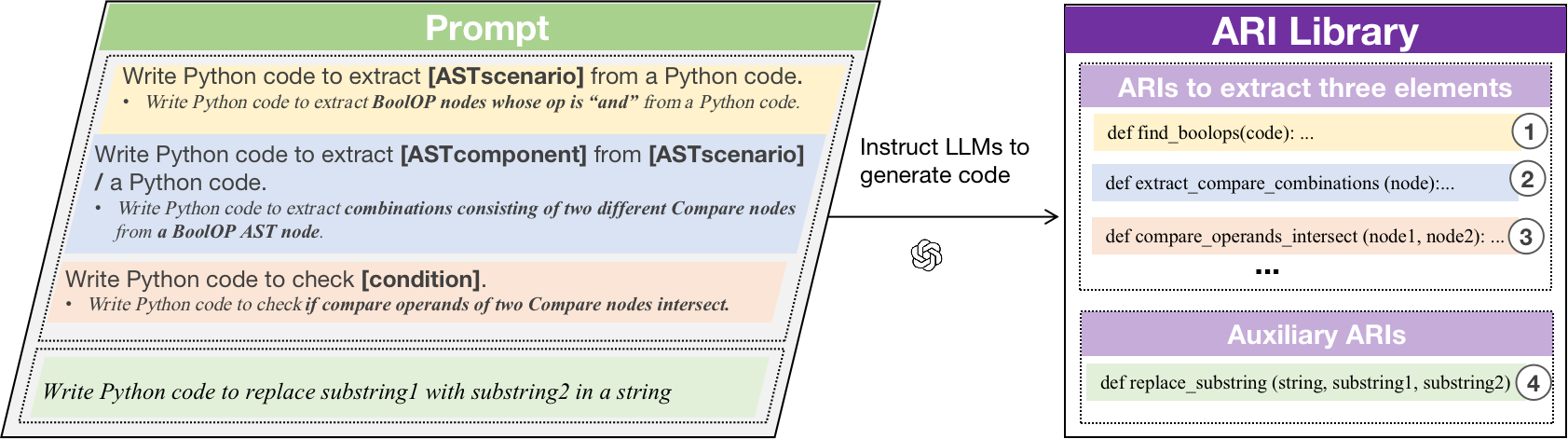}
    \caption{ARI library built by prompting LLMs to generate code}
    \label{fig:APILib} 
    \vspace{-0.3cm}
\end{figure}

The Analytic Rule Interface (ARI) library consists of ARIs to extract three elements in Table~\ref{tab:three_ele_idioms} and auxiliary ARIs. 
In contrast to directly relying on prompts to instruct LLMs in extracting the three essential elements from a given Python code, we employ LLMs to generate code to implement the required functionality. 
This approach addresses two key considerations. 
Firstly, within a project, it may have thousands of lines of code, and the non-idiomatic code is a small part of it. 
It is expensive for LLMs to handle so many codes and is difficult to make sure the non-idiomatic code is not missing and is correct from unrelated code in a given Python code. 
Secondly, by generating code through a single invocation of LLMs, we establish a reusable ARI library that can be leveraged consistently across different given Python code. 
Figure~\ref{fig:APILib} shows examples to prompt LLMs to generate code.

\noindent \textbf{Prompt LLMs to generate ARIs to extract three elements:} We first create three prompt templates for the three elements: ASTscenario, ASTcomponent and Condition. 
Then we instantiate the prompt templates with three elements of each Pythonic idiom from Table~\ref{tab:three_ele_idioms}. 
Finally, we instruct the LLM to generate code. 
Following the retrieval of the generated Python code, authors manually validate its correctness. 
Once verified, the code is cataloged as a reusable ARI, poised for application in subsequent any Python code~\footnote{We manually verify that all ARIs are correct}. 
This systematic approach ensures the reliability and reusability of the generated code for element extraction.

The template of ASTscenario is ``Write Python method code to extract [ASTscenario] from a Python code'', 
The [ASTscenario] is a placeholder which corresponds to the ASTscenario of a Pythonic idiom in Table~\ref{tab:three_ele_idioms}. 
For example, for the chain-comparison idiom in Figure~\ref{fig:APILib}, the template is instantiated into ``\textit{Write Python method code to extract BoolOP nodes whose op is ``and''}.'' 
When the prompt is input into the LLM, the LLM responds with an ARI called ``\textsf{extract\_and\_boolops(code)}''. 

The template of ASTcomponent is ``Write Python method code to extract [ASTcomponent] from [ASTscenario] / a Python code''. 
The [ASTcomponent] and [ASTscenario] are placeholders which corresponds to the ASTscenario and ASTcomponent of a Pythonic idiom in Table~\ref{tab:three_ele_idioms}. 
For example, for the chain-comparison in Figure~\ref{fig:APILib} and Table~\ref{tab:three_ele_idioms}, its ASTscenario is present, the template is instantiated into  ``\textit{Write Python method code to extract combinations consisting of two different Compare nodes from a BoolOP node}''. 
When the prompt is input into the LLM, the LLM responds with an ARI called ``\textsf{extract\_compare\_combinations(node)}''. 
For another example, for the list-comprehension, the ASTscenario is absent as shown in Table~\ref{tab:three_ele_idioms}, the the template is instantiated into  ``\textit{Write Python method code to extract For nodes from a Python code}''.  
When the prompt is input into the LLM, the LLM responds with an ARI called ``\textsf{extract\_for\_nodes(code)}''.

The template of Condition is ``Write Python method code to check [condition]''. 
The [condition] is a placeholder which corresponds to the Condition of a Pythonic idiom in Table~\ref{tab:three_ele_idioms}. 
For example, for the chain-comparison in Figure~\ref{fig:APILib}, the template is instantiated into  ``\textit{Write Python method code to check if compare operands of two Compare nodes intersect}''. 
When the prompt is input into the LLM, the LLM responds with an ARI called ``\textsf{compare\_operands\_intersect(node1, node2)}''. 
\noindent \textbf{Prompt LLMs to generate auxiliary ARIs:} 
For the auxiliary ARIs, it is a replace operation utilized in the idiomatization module. 
Since the replace operation is a simple task, we do not need to instruct LLMs to complete for each given Python code. 
And invocating the ARI can correctly and effectively replace substring1 with substring2 in a string. 
Specifically, we input the prompt ``\textit{Write Python method code to replace substring1 with substring2 in a string}'' into the LLM, the LLM responds with ``\textsf{replace\_substring(string, substring1, substring2)}'' whose body is ``\textsf{return string.replace(substring1, substring2)}''. 

\subsection{Extraction Module} ~\label{extract_module_with_APIs}
 \begin{figure}
\vspace{-0.2cm}
  \centering
   \includegraphics[width=5.3in]{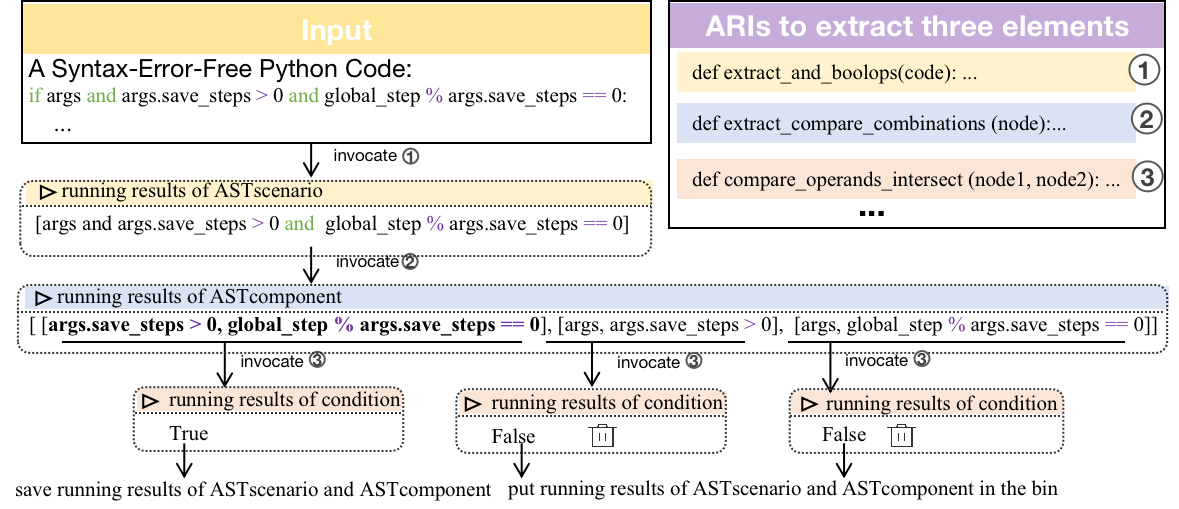} 
    \setlength{\abovecaptionskip}{2mm}
    \caption{Examples of extraction module}
    \label{fig:extract_module_example} 
    \vspace{-0.3cm}
\end{figure}
After we construct the knowledge base of three elements of non-idiomatic code of Pythonic idioms and ARIs to extract three elements. 
We call ARIs from ARI library in order ASTscenario, ASTcomponent and Condition. 
To elaborate, if the ASTscenario exists, we first extract ASTscenario from a Python code, and then extract the ASTcomponent from the ASTscenario followed by filtering out components that do not satisfy the condition if condition exists. 
If the ASTscenario does not exist, we directly extract the ASTcomponent from a Python code followed by filtering out ASTcomponent that does not satisfy the condition if the condition exists. 
For example, for the chain-comparison of Figure~\ref{fig:extract_module_example}, we first call ``\textsf{extract\_and\_boolops}'' ARI to get all BoolOP nodes from a Python code. 
Then, for each BoolOP node, we call ``\textsf{extract\_compare\_combinations}'' ARI to extract all two different Compare nodes from the BoolOP node. 
Finally, for each two Compare nodes, we call ``\textsf{compare\_operands\_intersect}'' ARI to filter out two Compare nodes without common comparison operands. 
For another example, for the list-comprehension, its ASTscenario is empty, we directly call ``\textsf{extract\_for\_nodes(code)}'' to extract all For nodes from a Python code, and then we call ``\textsf{has\_append(node)}'' filter out For nodes without ``\textsf{append}'' function call whose function name is the assigned variable of the Assign node. 


\subsection{Idiomatization Module
}

After extracting the code of ASTscenario and ASTcomponent, we can do the idiomatization task. 
Since the diversity and complexity of refactorable non-idiomatic code, it is not easy to complete the task by formulating rules.  
For example, consider code~\textcircled{1} Figure~\ref{motiv_example}. 
To refactor the code containing two continue statement with set-comprehension, we need to change ``\textsf{z is x}'' and ``\textsf{z not in df}'' into ``\textsf{z is not x}'' and ``\textsf{z in df}'', and then we use the ``\textsf{and}'' to connect the two conditions. 
Finally, we transform it with set-comprehension. 
Similarly, in the case of ~\textcircled{6} of Figure~\ref{motiv_example}, to refactor non-idiomatic code with chain-comparison, we need to reverse compare operands and determine whether need to change comparison operators based on the code semantic.
Fortunately, LLMs trained on large corpora have rich knowledge and huge
potential to complete complex tasks with natural language prompts~\cite{brown2020language, openai2023gpt4,feng2023prompting,peng2023generative,huang2023ai}.
Therefore, we write prompts to instruct LLMs to transform non-idiomatic code into idiomatic code for Pythonic idioms. 
To correctly complete the refactoring task, we design three steps including abstracting code, idiomatizing code and rewriting code. 

\begin{figure}
  \centering
    \setlength{\abovecaptionskip}{2mm}
    \includegraphics[width=5.4in]{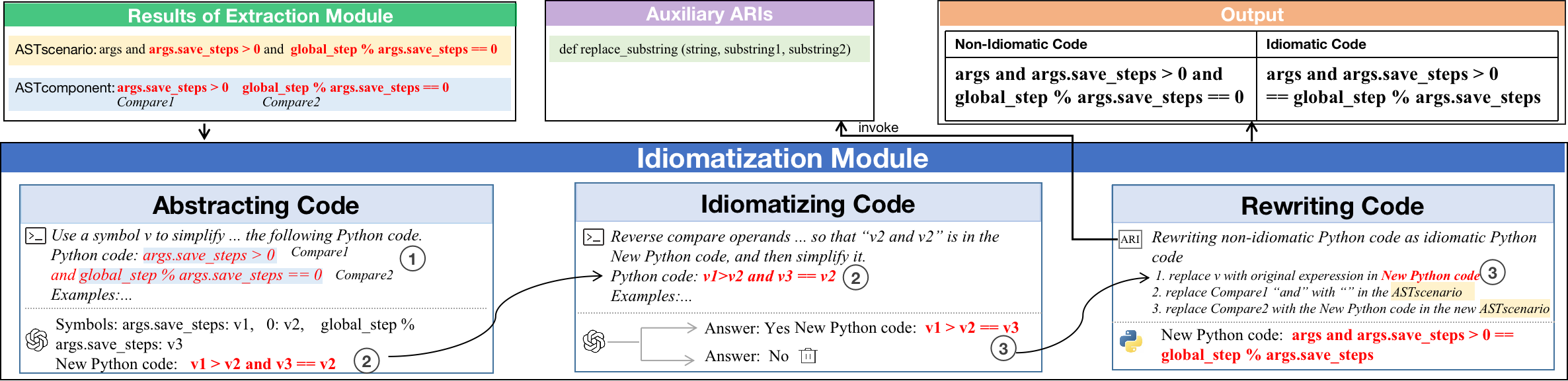}
    \caption{Examples of idiomatization module}
    \label{fig:idiomatize} 
    \vspace{-0.2cm}
\end{figure}

\subsubsection{Abstracting Code} ~\label{abstracting_code}
 The initial step involves the abstraction of the code which corresponds to ASTcomponent, wherein we keep related code snippets with Pythonic idioms while abstract unrelated code snippets with Pythonic idioms. 
 By distilling the code to its core elements, we enhance the clarity and simplicity of the subsequent idiomatization process. 
To determine the abstract form for each Pythonic idiom, we invite three external workers with more than five years Python programming experience. 
We first introduce Pythonic idioms to make sure they are familiar with the idioms. 
Each worker is then asked to independently review the non-idiomatic code dataset of Pythonic idioms from previous researches~\cite{zhang2022making, Zen_Your_Python}, and then writes a description of abstracting code and provides three examples of original non-idiomatic code and abstract non-idiomatic code. 
Then two authors discuss their results and give the final description of abstracting code (\textit{<prompt>}) and three examples of original non-idiomatic code and the corresponding abstract non-idiomatic code for each Pythonic idiom (\textit{<examples>}). 

For the abstracting code for each Pythonic idiom, if the prompt has a specified object to abstract, we use ARIs from Auxiliary ARIs in Section~\ref{APIs} to replace a given string with another given string. 
Otherwise, we use prompt to instruct LLMs to complete the task. 
For example, for the star-in-func-call, the extracted non-idiomatic code is ``\textsf{feat.shape[-2], feat.shape[-1]}'' and the abstracted expression is a specified object (``\textsf{feat.shape}'').
Therefore, we invoke ``\textsf{replace\_substring}'' from auxiliary ARIs to replace  ``\textsf{feat.shape}''with ``\textsf{v}'', so the abstract code is ``\textsf{v[-2], v[-1]}''. 
For another example in the Abstracting Code of Figure~\ref{fig:idiomatize}, 
it does not have a specified object to abstract, 
We use prompt ``\textit{Use a symbol v to simplify each comparison operand within the following Python code. 
The same comparison operand is represented by the same symbol.}'' to abstract represent the code ``\textsf{args.save\_steps > 0 and global\_step \% args.save\_steps == 0}''. 
The LLM responds with a symbol mapping and the abstract Python code ``\textsf{v1 > v2 and v3 == v2}'', facilitating subsequent idiomatizing process. 




\subsubsection{Idiomatizing code} Building upon the abstracted Python code representation, this step focuses on the actual idiomatization of the code. 
To determine the prompt of idomatizing code for each Pythonic idiom, following the similar process~\ref{abstracting_code}, we invite the same three external works to further independently write a description to idiomatize the abstract code for each Pythonic idiom and provide  examples with the abstract code and the corresponding idiomatic code for each Pythonic idiom. 
And then two authors discuss their results to give the final description of idiomatizing code (\textit{<prompt>}), and the abstract code and the corresponding idiomatic code for each Pythonic idiom (\textit{<examples>}). 

For example, for the chain-comparison in the Idiomatizating Code of Figure~\ref{fig:idiomatize}, we use prompt ``\textit{Reverse compare operands of the first comparison operation, the second comparison, or the first and the second comparison operations so that ``\textsf{v2 and v2}'' is in the new Python code, and then simplify it}'' to idiomatize the abstract Python code: ``\textsf{v1 > v2 and v3 == v2}''. 
The LLM responds with Yes and the abstract idiomatic Python code ``\textsf{v1 > v2 == v3}''. 
For another example, for the abstract code of chain-comparison  ``\textsf{v1 in v2 and v3 in v2}'', The LLM responds with No because reversing compare operands is invalid for the ``\textsf{in}'' operator that can change the code semantic. 



\subsubsection{Rewriting code} 
Following the acquisition of abstract idiomatic code from the idiomatization process, the final step involves rewriting the non-idiomatic code. 
It is achieved through the application of the ``\textsf{replace}'' ARI sourced from the Auxiliary ARIs. 
The ``\textsf{replace}'' operation serves a dual purpose: it facilitates the restoration of the abstract idiomatic code and allows for direct code rewriting by replacing the ASTcomponent with the idiomatic code in the ASTscenario, if ASTscenario exists. 
Therefore, it may invoke ``\textsf{replace}'' several times. 
To determine the process of invoking ``\textsf{replace}'', following the similiar process in Section~\ref{abstracting_code}, we invite the same three external workers to further independently summarize steps to use ``replace'' to complete rewriting non-idiomatic code into idiomatic code. 
And then two authors discuss their results to give the final steps of rewriting code for each Pythonic idiom. 

For example, for the Rewriting Code process illustrated in Figure~\ref{fig:idiomatize}, 
we first replace abstract symbols with their corresponding original expressions within the abstract idiomatic code: ``\textsf{v1 > v2 ==v3}'', yielding the genuine idiomatic code.  
Subsequently, we replace the Compare1 and `and' with an empty string in the ASTscenario code, yielding the new ASTscenario code.
Finally, we replace the Compare2 with the genuine idiomatic code in the new ASTscenario code, yielding a final idiomatic code:  ``\textsf{args and args.save\_steps > 0 == global\_step \% args.save\_steps}''. 
The rewriting step ensures the precise transformation of non-idiomatic code into its idiomatic counterpart.


%% file: result.tex
To evaluate our approach, we study two research questions:

\begin{enumerate}[fullwidth,itemindent=0em,leftmargin = 0pt]
	\item[\textbf{RQ1 (Effectiveness):}] What is the effectiveness of our approach in refactoring non-idiomatic Python code into idiomatic Python code with nine Pythonic idioms?
	\item[\textbf{RQ2 (Scalability):}] Can our approach be effectively extended to new Pythonic idioms? 
\end{enumerate}

\subsection{RQ1: Effectiveness of Refactoring Non-Idiomatic Python Code with Nine Pythonic Idioms}\label{rq1}

\subsubsection{Motivation} RIdiom~\cite{zhang2022making} can automatically refactor non-idiomatic code into idiomatic code with nine Pythonic idioms by formulating detection and refactoring rules, but it causes huge human investment in formulating rules.  
The current success of ChatGPT~\cite{intro_chat_gpt} demonstrates remarkable ability of LLMs to comprehend human prompts and complete the corresponding tasks. 
Therefore, we are interested in understanding the performance of our approach based on LLMs.

\subsubsection{Approach}\label{rq1_approa} To clarify the effectiveness of our approach, we perform effectiveness comparison by calculating metrics on a dataset with our approach and the state-of-the-art baselines.


\noindent \textbf{DataSet.} 
To evaluate the effectiveness of our approach, it is fundamental to have a correct and complete benchmark of code refactorings consisting of code pairs <non-idiomatic Python code, idiomatic Python code>. 
Manually constructing a complete and correct benchmark is unrealistic because code refactoring involves a lot of time and manpower, and inevitably comes with personal bias. 
Recently, RIdiom~\cite{zhang2023ridiom} can automatically refactor non-idiomatic Python code with nine Pythonic idioms, which provides a good starting point. 
It provides code pairs within crawled methods for each Pythonic idiom. 
Considering the effort of manual verification, we randomly sample methods with a confidence level of 95\% and a confidence interval of 5 from RIdiom~\cite{zhang2023ridiom} for each Pythonic idiom. 
Then, to ensure the completeness of the benchmark, we run RIdiom~\cite{zhang2023ridiom}, our approach and Prompt-LLM on the sample methods for nine Pythonic idioms to collect code pairs <non-idiomatic Python code, idiomatic Python code>.
To validate the correctness, we invite 18 external workers with more than five years Python programming experience. 
We divide them into 9 groups, and each group of two workers independently checks the correctness of the code pairs for an idiom. 
The Cohen's Kappa values~\cite{viera2005understanding} of nine groups for their annotation results all exceeded 0.75 (substantial agreement). 
Finally, the two authors and external workers discuss and resolve the inconsistencies and ensure the correctness of the benchmark for nine Pythonic idioms. 
Table~\ref{benchmark} shows the benchmark of nine Pythonic idioms. 
The \textit{Method} column represents the number of sampled methods for each Pythonic idiom, and the \textit{Code Pair} column represents the number of code pairs <non-idiomatic code, idiomatic code> for each Pythonic idiom. 
Since sampled methods for each Pythonic idiom are from methods of RIdiom, it is reasonable to expect that the number of code pairs is greater than the number of the corresponding methods for each Pythonic idiom. 
\begin{table}
\footnotesize

\caption{Benchmark of Nine Pythonic Idioms}
  \label{benchmark}
\vspace{-0.4cm}
\centering
\begin{tabular}{|l|c|c|}
\hline
Idiom & Method & Code Pair \\ \hline
list-comprehension & 389 & 512
\\ \hline
set-comprehension & 305 & 361 \\ \hline
dict-comprehension & 370 & 448
\\ \hline
chain-comparison & 391 & 574 \\ \hline
truth-test & 394 & 610 \\ \hline
loop-else & 319 & 360 \\ \hline
assign-multi-targets & 395 & 729 \\ \hline
for-multi-targets & 370 
& 463 
\\ \hline
star-in-func-call & 381 & 621 \\ \hline
Total & 3311 & 4678
\\ \hline
\end{tabular}
\vspace{-0.4cm}
\end{table}

\noindent \textbf{Baselines.} 
We compare our approach with two baselines. 
The first baseline is RIdiom that is an approach based on rules proposed by Zhang et al.~\cite{zhang2022making}. 
It detects and refactors non-idiomatic Python code into the corresponding idiomatic Python code with nine Pythonic idioms by manually formulating detection rules and refactoring steps.
The second (Prompt-LLM) is to directly call the LLM to find code pairs for any given method code for each Pythonic idiom to illustrate the capability of LLM and the strengths of our proposed approach using LLM. 
For fairness of comparison, similar to process in Section~\ref{abstracting_code}, we invite three external works to independently write a prompt and provide three examples of a Python code and the corresponding code pairs. 
Then two authors discuss their results and give the final prompt and three examples for each Pythonic idiom. 
Examples are shown in Prompt-LLM column of Figure~\ref{motiv_example}. 
Since the success of ChatGPT, the LLM our approach and baselines use are state-of-the-art of GPT-3.5-turbo~\cite{chat_gpt}. 
And we set temperature to 0 to make the outputs mostly deterministic.

\noindent \textbf{Metrics.} 
By referring metrics using by previous researches on code refactorings~\cite{zhang2022making,tsantalis2018accurate,dilhara2023pyevolve,silva2017refdiff}, we use four metrics accuracy, F1-score, precision and recall. 
To calculate the four metrics, we need to define true positives, false positives and false negatives which are represented as $TP$, $FP$ and $FN$, respectively.
We define a $TP$ as a code pair detected by an approach is in the benchmark. 
We define a $FP$ as a code pair detected by an approach is not in the benchmark.
We define a $FN$ as a code pair of the benchmark is not determined by an approach.
The accuracy and F1-score represents the overall performance. 
We calculate accuracy, F1-score, precision and recall as follows: 
$$Precision = \frac{TP}{TP+FP},\ \ \ Recall = \frac{TP}{TP+FN}$$
$$F1 = \frac{2*P*R}{P+R},\ \ \ Accuracy = \frac{TP}{TP+FP+FN}$$

\subsubsection{Result}

\begin{figure}
\vspace{-0.2cm}
  \centering
   \includegraphics[width=5.3in]{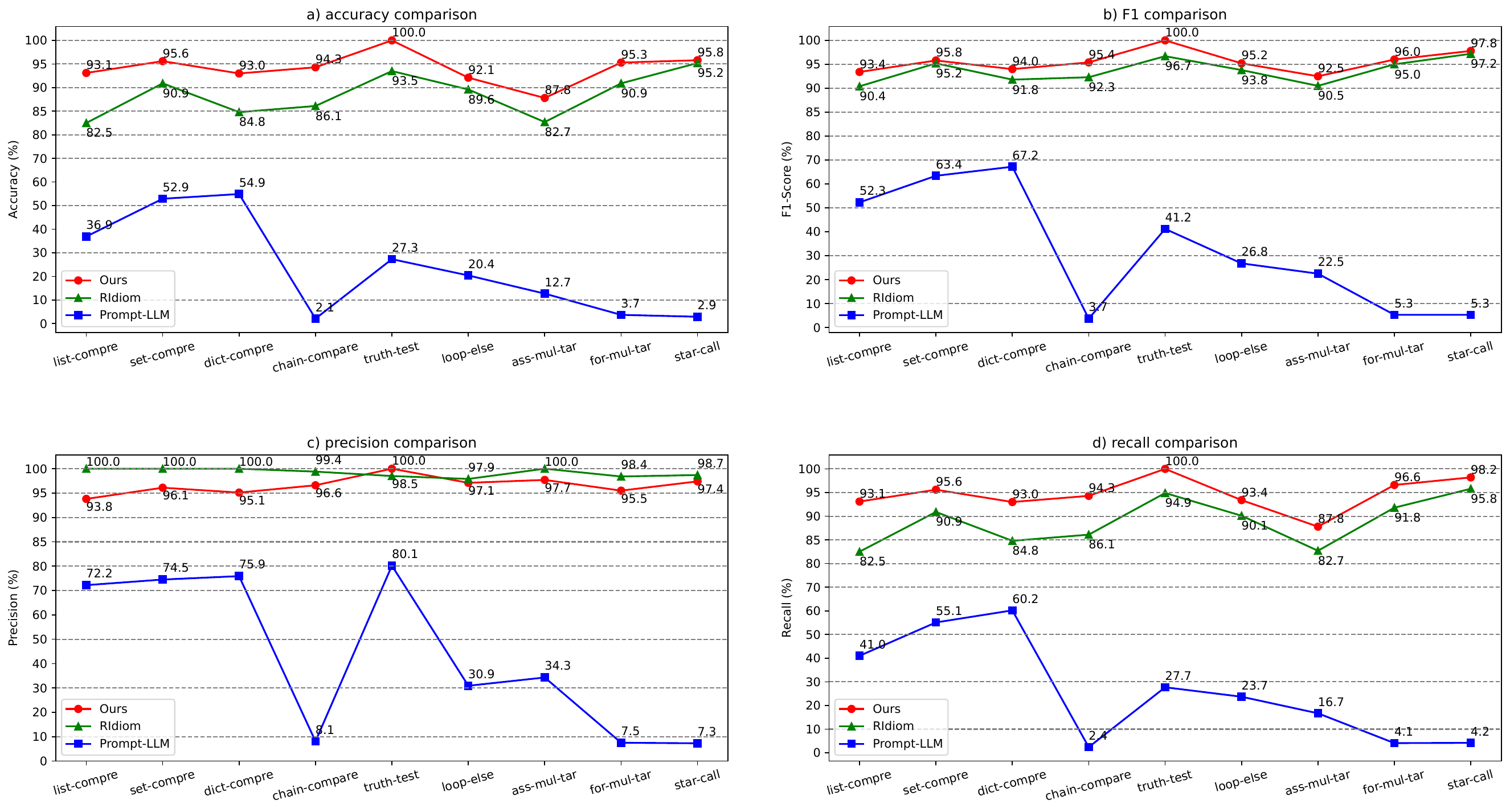}
 \setlength{\abovecaptionskip}{1mm}
 
    \caption{Scatter plot with straight lines of accuracy, F1-score, precision and recall of three approaches for nine Pythonic idioms}
    \label{correctness_1} 
    \vspace{-0.5cm}
\end{figure}

Figure~\ref{correctness_1} presents the accuracy, F1-score, precision and recall of our approach, RIdiom, Prompt-LLM for nine Pythonic idioms. 
\noindent \textbf{Comparison with RIdiom.} In comparison to RIdiom, our approach consistently outperforms in three metrics: accuracy, F1-score, and recall across all Pythonic idioms. 
Notably, our approach exhibits a distinct advantage in both recall and accuracy over RIdiom. 
For eight Pythonic idioms, the accuracy and the recall exceed 90\%. 
While our approach registers slightly below 90\% in both recall and accuracy for the assign-multi-targets idiom, it is obviously close to 90\% (87.8\%). 
In contrast, RIdiom falls short of 90\% in accuracy for five Pythonic idioms (list-comprehension, dict-comprehension, chain-comparison, loop-else, and assign-multi-targets), and four of these idioms (list-comprehension, dict-comprehension, chain-comparison, and assign-multi-targets) also have recall results below 90\%. 
For the list-comprehension, the disparity in recall and accuracy between our approach and RIdiom exceeds 10\%. 
For the other four idioms (dict-comprehension, chain-comparison, truth-test, and assign-multi-targets), these differences surpass 5\%. 
For the remaining four idioms (set-comprehension, loop-else, for-multi-targets, and star-in-func-call), while the differences are relatively smaller, our approach maintains a consistent edge over RIdiom.  

For the precision, although RIdiom exhibits a slight advantage in precision (with differences ranging from 0.8\% to 6.2\%) for eight of the idioms, our approach consistently achieves over 93.8\% precision for each idiom. 
Moreover, our approach surpasses RIdiom for the truth-test idiom. 
It is important to note that F1-score strikes a balance between precision and recall, and in this regard, our approach consistently outperforms RIdiom across all Pythonic idioms. 
 The comprehensive analysis underscores the notable advantages of our approach over RIdiom for the task of refactoring non-idiomatic code with Pythonic idioms, and can make up for the shortcomings of RIdiom in recall. 

\noindent \textbf{Comparison with Prompt-LLM.} Compared to our approach, Prompt-LLM consistently exhibits the lowest performance across four metrics for each Pythonic idiom. 
The disparities between our approach and Prompt-LLM in terms of accuracy, F1-score, precision, and recall for the nine Pythonic idioms are substantial, ranging from 38.1\% to 92.9\%, 26.8\% to 92.5\%, 19.9\% to 90.1\%, and 32.8\% to 94\%, respectively.

Compared to Prompt-LLM, our approach maintains stable and commendable performance across all idioms, with each metric surpassing the 90\% threshold for eight of them. 
Even in the case of assign-multi-targets, where both accuracy and recall fall slightly below 90\%, they are still notably close at 90\% (87.8\%). 
In contrast, Prompt-LLM fails to achieve a metric score of 90\% for any of the idioms. 

Prompt-LLM demonstrates poor performance and exhibits significant variability across different Pythonic idioms.   
For list/set/dict-comprehension, Prompt-LLM displays relatively better performance across all metrics, exceeding 35\%. 
This underscores LLMs' enhanced proficiency in handling the three idioms. 
For truth-test, Prompt-LLM exhibits superior precision (80.1\%) compared to recall (27.7\%), indicating a higher likelihood of missing refactorable non-idiomatic code instances associated with the truth-test idiom. 
For loop-else and assign-multi-targets, Prompt-LLM's performance remains limited across all metrics, falling below 35\%. 
Furthermore, for the remaining three idioms (chain-comparison, for-multi-targets, and star-in-func-call), Prompt-LLM's performance is markedly poorer, registering below 10\% on all metrics. 
It suggests that Prompt-LLM struggles to effectively detect and refactor non-idiomatic code associated with the three idioms.
Therefore, our approach significantly enhances the capabilities of LLMs, consistently demonstrating stable and commendable performance across all metrics. 

\begin{figure}
  \centering
  \includegraphics[width=5.4in]{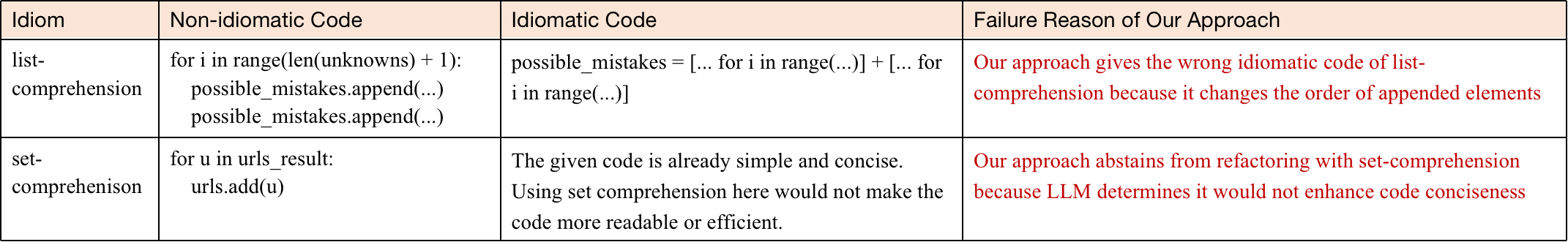}
 \setlength{\abovecaptionskip}{2mm}
    \caption{The examples that our approach wrongly refactors or refrains from refactoring}
    \label{fig:failure_example_our} 
    \vspace{-0.3cm}
\end{figure}

\noindent \textbf{Failure analysis of our approach.} 
For the code pairs in the benchmarks that our approach does not find or wrongly refactor non-idiomatic Python code with nine Pythonic idioms, we summarize two reasons as follows:


    
   (1) LLMs may produce suboptimal results when refactoring is too complicated. 
    While LLMs have achieved success, it is reasonable that LLMs cannot handle all situations. 
    For example, for the first example of Figure~\ref{fig:failure_example_our}, the non-idiomatic code appends two different elements to the list ``\textsf{possible\_mistakes}'' in each iteration of the for statement. 
    Our approach finally gives the idiomatic code by concatenating two lists which independently appending two elements. 
    The idiomatic code is wrong because the order of elements is different from the non-idiomatic code. 
    Although the non-idiomatic code can be refactored with list-comprehension, the idiomatic code needs other statements to adjust the order of elements.
    For another example, for the non-idiomatic code ``\textsf{gnn\_layer(..., n\_points[idx1], n\_points[idx2])}'', it is non-refactorable code with star-in-func-call because the ``idx1, idx2'' is not an arithmetic sequence. 
    However, our approach mistakenly assumes the subscript sequence of  ``\textsf{\_points[idx1], n\_points[idx2]}'' is an arithmetic sequence and refactor it into ``\textsf{gnn\_layer(..., *n\_points[idx1:idx2 + 1])}'' with star-in-func-call. 
    
    (2) LLMs may refrain from refactoring when benefits in idiomatic code appear limited: While LLM demonstrates proficiency in refactoring Python code using specific Pythonic idioms, there are instances where it abstains from doing so. 
    For example, for the second example of Figure~\ref{fig:failure_example_our}, the non-idiomatic code ``\textsf{for u in urls\_results: urls.add(u)}'' actually can be refactored with set-comprehension. 
    However, LLM responds with \textit{``The given code is already simple and concise. 
    Using set comprehension here would not make the code more readable or efficient.
''}. 
LLM refuses to refactor it with set-comprehension because it may determine that applying set-comprehension would not notably enhance code conciseness. 
    For another example, for the non-idiomatic code ``\textsf{slice2[axis] = slice(None, -1); slice1 = tuple(slice1)}'', LLM responds with \textit{``The given code cannot be refactored with one assign statement as the variables being assigned are not of the same type''}.
    The LLMs refuses to refactor because it thinks ``\textsf{slice2[axis]}'' and ``\textsf{slice1}'' are not the same type.  
    
\vspace{1.5mm}
\noindent\fbox{\begin{minipage}{5.4in} \emph{Our approach can achieve best accuracy, F1-score and recall compared to other baselines. 
Although our approach performs slightly worse than RIdiom on precision, our approach can complement RIdiom on recall.} \end{minipage}}

\subsection{RQ2: Scalability of Our Approach}\label{usefulness}

\subsubsection{Motivation} Although RIdiom can achieve good result on nine Pythonic idiom, it cannot handle new Pythonic idioms and is difficult to extend to new Pythonic idioms because RIdiom needs manually formulate extracting and refactoring rules. 
So we are interested in whether our approach can be effectively extended to new Pythonic idioms that RIdiom cannot handle.

\subsubsection{Approach} Following the approach in Section~\ref{rq1_approa}, we present baselines, dataset and metrics. 
Baselines and metrics are the same as metrics and baselines in Section~\ref{rq1_approa}. 
Particularly, since RIdiom cannot handle the new four Pythonic idioms, the baseline is excluded. 
The dataset is detailed as follows:
\begin{table}
\footnotesize

\caption{Benchmark of Four New Pythonic Idioms}
  \label{rq2_res}
\vspace{-0.2cm}
\centering
\begin{tabular}{|l|c|c|}
\hline
Idiom & Methods & Code Pairs \\\hline

with & 600 & 64 \\ \hline
enumerate & 600 & 565 \\ \hline
chain-assign-same-value & 600 & 156 \\ \hline
fstring  & 600 & 223
\\ \hline
Total & 600 & 1008 \\ \hline
\end{tabular}
\vspace{-0.1cm}
\end{table}
\begin{figure}
\vspace{-0.1cm}
  \centering
   \includegraphics[width=5.4in]{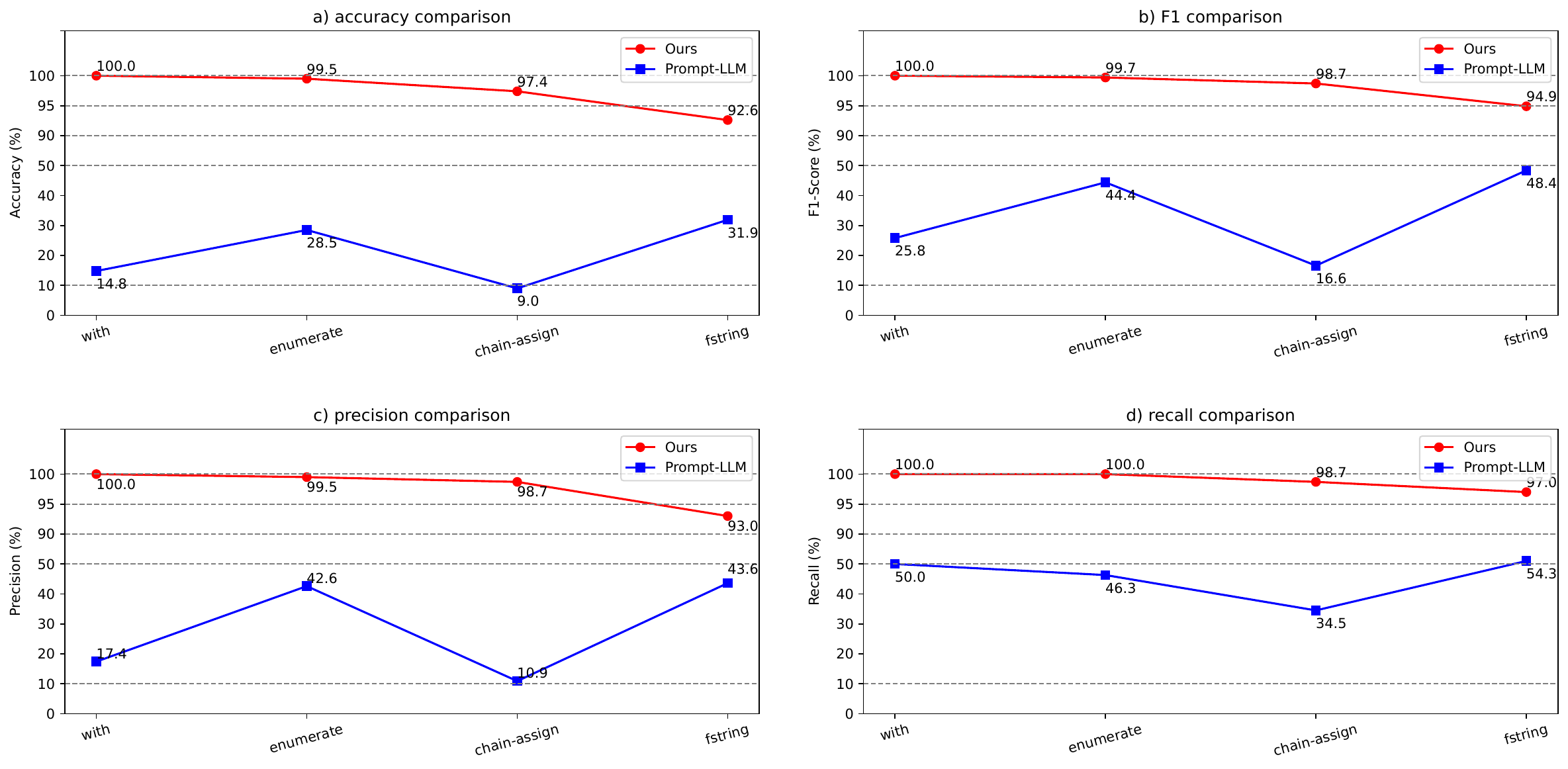}
\setlength{\abovecaptionskip}{1mm}
    \caption{Scatter plot with straight lines of accuracy, F1-score, precision and recall of two approaches for four new Pythonic idioms}
    \label{correctness_2} 
     \vspace{-0.5cm}
\end{figure}
\noindent \textbf{DataSet.} 
As Section~\ref{idiom_library} illustrates, there are four new Pythonic idioms (with, enumerate, fstring and chain-assign-same-value) that RIdiom cannot handle. 
Considering that we need a certain number of code pairs (<non-idiomatic Python code, idiomatic Python code>) to evaluate our approach, but manually verifying code pairs is time consuming and the current GPT-3.5 is not free, we randomly sample 600 Python methods from crawled methods by RIdiom~\cite{zhang2022making} that is a tool to refactor non-idiomatic code into idiomatic code with nine Pythonic idioms. 
Following the similar method in Section~\ref{rq1}, to ensure the completeness of dataset as much as possible, we run our approach and Prompt-LLM on the sampled Python methods to collect code pairs <non-idiomatic Python code, idiomatic Python code>. 
To validate the correctness, we invite 8 external workers with more than five years Python programming experience. 
We divide them into four groups, and each group of two workers independently checks the correctness of the code pairs for an idiom. 
The Cohen's Kappa values~\cite{viera2005understanding} of four groups for their annotation results all exceeded 0.75 (substantical agreement). 
Finally, the two authors and external workers discuss and resolve the inconsistencies and ensure the correctness of the benchmark for the new four Pythonic idioms. 
Table~\ref{rq2_res} shows the benchmark of new four Pythonic idioms. 
The \textit{Method} column represents the number of sampled methods for each Pythonic idiom, and the \textit{Code Pair} column represents the number of code pairs <non-idiomatic code, idiomatic code> for each Pythonic idiom. 
The number of code pairs is less than the number of methods is reasonable because not each method may contain non-idiomatic code. 
For example, in the case of the with idiom, its non-idiomatic code should include file opening operations. 
However, not all methods necessarily involve file opening. Given that the with idiom is widely adopted by Python users~\cite{sakulniwat2019visualizing}, the count is relatively low (64). 
Conversely, for the enumerate idiom, its non-idiomatic code typically corresponds to a for statement, which is more prevalent in code. 
As a result, the count is relatively higher (565).

\subsubsection{Result} 
Figure~\ref{correctness_2} presents the accuracy, F1-score, precision and recall of our approach, and Prompt-LLM for new four Pythonic idioms. 
For each Pythonic idioms, the metrics all are above 90\%. 
And the accuracy, F1-score, precision and recall are above 95\% for three idioms (with, enumerate and chain-assign-same-value). 
Compared to our approach, Prompt-LLM consistently exhibits poor performance across four metrics for each Pythonic idiom. 
The disparities between our approach and Prompt-LLM in terms of accuracy, F1-score, precision, and recall for the four Pythonic idioms are substantial, ranging from 60.7\% to 88.4\%, 46.5\% to 82.1\%, 
49.4\% to 87.8\%, and 42.7\% to 64.2\%, respectively.

\vspace{1mm}
\noindent\fbox{\begin{minipage}{5.4in} \emph{ The high accuracy, F1-score, precison and recall of code refactorings of the four new Pythonic idioms illustrate that our approach can be effectively extended to other Pythonic idioms. 
} \end{minipage}}


%% file: discuss.tex

\subsection{Implications} 
Our hybrid approach in Section~\ref{method} demonstrates excellent performance in Section~\ref{result}, we now delve into the scalability of our approach and future work for researchers. 
Currently, our approach only supports syntactically correct Python code, as it requires parsing the code into an Abstract Syntax Tree (AST). 
However, handling Python code with syntax errors on the hybrid framework is feasible. 
One solution is to incorporate a syntax-fixing module to rectify syntax errors in the Python code before it input into extraction module in  Section~\ref{extract_module_with_APIs}. 
Another approach is to introduce an alternative method within the extraction module in Section~\ref{extract_module_with_APIs}. 
Specifically, prompting LLMs to extract three elements or prompting LLMs to generate AST for Python code instead of invoking APIs, which can be effective when the syntax error cannot be fixed. 

Prior studies~\cite{faster_slower,alexandru2018usage, Zen_Your_Python,zhang2022making} have shown that employing Pythonic idioms may yield benefits, such as code conciseness and improved performance, but it can also have drawbacks, including potential impacts on code readability and performance degradation. 
Our current focus is solely on refactoring non-idiomatic code with Pythonic idioms. 
The results in Section~\ref{result} and Figure~\ref{fig:failure_example_our} highlight that LLMs may abstain from refactoring when benefits from idiomatic code are limited. 
This observation prompts researchers to consider using LLMs combined with knowledge base to generate comments explaining the positive and negative effects of refactoring non-idiomatic code with Pythonic idioms. 
This can enhance Python users' understanding and effective utilization of Pythonic idioms.

Furthermore, our approach combines ARIs and prompts based on LLMs to refactor non-idiomatic code with Pythonic idioms. 
While current Language Models (e.g., ChatGPT~\cite{openai2023gpt4}) are not freely available, there has been a recent emergence of free alternatives (e.g., Llama 2~\cite{touvron2023llama}). 
Over time, it is plausible that more LLMs may become more accessible, potentially even free of charge. This accessibility could assist developers in effectively refactoring code using Pythonic idioms to alleviate the limitation of rule-based approach.

\subsection{Threats to Validity}
\textbf{Internal Validity:}
The one internal threat is inaccuracy when evaluating the correctness of benchmark in Section~\ref{result}. 
For each Pythonic idiom, we invite two external workers with more than five Python programming experience. 
And two authors and external workes discuss to  resolve the inconsistencies to ensure the correctness of the benchmark.

\noindent \textbf{External Validity:} 
One external threat is that our approach is limited to thirteen Pythonic idioms. 
RIdiom~\cite{zhang2023ridiom} can only handle nine Pythonic idioms, our approach contains four new Pythonic idioms which validate the scalability of our approach. 
When there are more Pythonic idioms, developers can determine the three elements of Pythonic idioms to extend to more Pythonic idioms. 
In the future, we will automatically mine undetected Pythonic idioms and automatically determine three elements of Pythonic idioms. 
The other external threat is the representative of the benchmark selected to evaluate our approach. 
To mitigate this threat, our experimented methods of benchmark are from previous research~\cite{zhang2023ridiom}, representing an unbiased benchmark for our research.

%% file: related.tex
\noindent \textbf{Studies on Pythonic idioms.} 
Pythonic idioms are highly valued by researchers, there are several studies~\cite{Merchante2017FromPT,zhang2023ridiom,alexandru2018usage,Zen_Your_Python,knupp2013writing,hettinger2013transforming,localize_idiomatic_code,python3,deidiom} to mine Pythonic idioms or help Python users use Pythonic idioms better.
Alexandru et al.~\cite{alexandru2018usage} and Farooq et al.~\cite{Zen_Your_Python} conducted an independent literature review to create a catalogue of Pythonic idioms. 
There are 27 detectable Pythonic idioms involving built-in methods, APIs and syntax. 
Phan-udom et al.~\cite{phan2020teddy} first collected 58 non-idiomatic code instances and 55 idiomatic. 
Subsequently, they provided Pythonic code examples akin to code in projects of developer. 
Sakulniwat et al.~\cite{sakulniwat2019visualizing} proposed a technique to visualize and understand the usage of the with Pythonic idiom and found developers tend to adopt the idiomatic code over time. 
Dilhara et al.~\cite{dilhara2023pyevolve} mined frequent code changes in Python ML systems and found some of involving Pythonic idioms.  
RIdiom~\cite{zhang2023ridiom} first identified nine Pythonic idioms by comparing the syntax difference between Python and Java and then designed detection and refactoring rules to automatically refactor non-idiomatic code into idiomatic code with the nine Pythonic idioms. 
Leelaprute et al.~\cite{leelaprute2022does} analyzed the performance of Python features (e.g., collections.defaultdict and lambda) and two Pythonic idioms (list/dict comprehension) with different input sizes. 
Zhang et al.~\cite{faster_slower} conducted a large-scale empirical study for nine Pythonic idioms by creating a synthetic dataset and real-project dataset in RIdiom~\cite{zhang2023ridiom}. 
They found Pythonic idioms do not always result in performance speedup and can cause degraded performance. 
Zhang et al.~\cite{deidiom} explored the challenges in comprehending Pythonic idioms, their conciseness, and the potential impact of comprehension challenges on code. 
They developed the DeIdiom tool to interpret these idioms into equivalent non-idiomatic code, facilitating developers in comprehending and effectively leveraging Pythonic idioms. 
In this work, we focus one refactoring non-idiomatic code with Pythonic idioms. 
Unlike RIdiom~\cite{zhang2023ridiom}, we do not rely on a rule-based approach; instead, we employ a hybrid approach utilizing APIs and prompts based on LLMs. 
Besides, we extend our approach to four new Pythonic idioms, which further verify the scalability of our approach.


\noindent \textbf{Studies on LLMs.} 
Building on the achievements of Language Models (LLMs) like GPT-3~\cite{brown2020language} and GPT-4~\cite{openai2023gpt4} in the field of Natural Language Processing~\cite{chen2022codet,unknown,AI_Chains_Wu,CodaMosa,Yang2022Re3GL,wei2023chainofthought,wu2022ai,code_generation_exception}, researchers are now delving into their potential applications in software engineering~\cite{huang2023ai,feng2023prompting,peng2023generative,codex,Jigsaw_jain,Prompt_Tune_qinghuang,VulRepair,zheliu_fill_blank,Li_2022,dilhara2024unprecedented}. 
Huang et al.~\cite{huang2023ai} introduced a chain-of-thought approach based on LLMs, comprising four steps: extracting structure hierarchy, isolating nested code blocks, generating CFGs for these nested blocks, and amalgamating all CFGs. 
This approach surpasses existing CFG tools in terms of both node and edge coverage, particularly for incomplete or erroneous code. 
Feng et al.~\cite{feng2023prompting} proposed a two-phase approach utilizing a chain-of-thought prompt to guide LLMs in extracting S2R entities, followed by matching these entities with GUI states to replicate the bug reproduction steps. This demonstrates that instructing LLMs through prompts can effectively achieve bug replay. 
Peng et al.~\cite{peng2023generative} presented TYPEGEN, which generates prompts incorporating domain knowledge, then feeds them into LLMs for type prediction. 
This approach uses few annotated examples to achieve superior performance compared to rule-based type inference approaches. 
In this work, we focus on making Python code idiomatic with Pythonic idioms. 
 Unlike previous approaches that exclusively design prompts for instructing LLMs to generate code or complete tasks, we propose a hybrid approach using APIs generated by LLMs to extract three elements, along with prompts to guide LLMs in performing the idiomatization task. 
 It demonstrates that LLMs and rule-based approaches complement each other which can assist researchers to solve software engineering tasks better.

\vspace{-1.3mm}